IAC-24,B3,IP,26,x83991

# Testing and Validation of Innovative Extended Reality Technologies for Astronaut Training in a Partial-Gravity Parabolic Flight Campaign

**Florian Saling[a]\*, Andrea E. M. Casini[a], Andreas Treuer[a], Martial Costantini[a], Leonie Bensch[a], Tommy Nilsson[b], Lionel Ferra[b]**

[a] *German Aerospace Center (DLR), Space Operations and Astronaut Training, Linder Höhe, 51147 Cologne, Germany*
[b] *European Space Agency (ESA), European Astronaut Centre (EAC), Linder Höhe, 51147 Cologne, Germany*
\* Corresponding Author: florian.saling@dlr.de

**Abstract**

The use of eXtended Reality (XR) technologies in the space domain has increased significantly over the past few years as it can offer many advantages when simulating complex and challenging environments. Space agencies are currently using these disruptive tools to train astronauts for Extravehicular Activities (EVAs), to test equipment and procedures, and to assess spacecraft and hardware designs.

With the Moon being the current focus of the next generation of space exploration missions, simulating its harsh environment is one of the key areas where XR can be applied, particularly for astronaut training. Peculiar lunar lighting conditions in combination with reduced gravity levels will highly impact human locomotion especially for movements such as walking, jumping, and running. In order to execute operations on the lunar surface and to safely live on the Moon for an extended period of time, innovative training methodologies and tools such as XR are becoming paramount to perform pre-mission validation and certification.

To further increase the maturity level of the associated XR technologies, the experts of the European Astronaut Centre (EAC) in Cologne, Germany, have developed multiple experiments that have been conducted during a partial-gravity parabolic flight campaign. The first experiment examines how well different commercial-of-the-shelf and custom-made XR headsets perform in lunar gravity while also evaluating how users can grasp and carry different payload mockups during a simulated lunar EVA. The second experiment uses multiple motion tracking cameras to accurately capture human locomotion in partial gravity conditions and to further improve existing locomotion models.

This research work presents the findings of the experiments aimed at exploring the integration of XR technology and parabolic flight activities for astronaut training. In addition, the study aims to consolidate these findings into a set of guidelines that can assist future researchers who wish to incorporate XR technology into lunar training and preparation activities, including the use of such XR tools during long duration missions.

**Acronyms/Abbreviations**

CNES: Centre National d'Etudes Spatiales (French National Centre for Space Studies)
CONOPS: Concept of Operations
COTS: Commercial-Off-The-Shelf
DLR: Deutsches Zentrum für Luft- und Raumfahrt (German Aerospace Center)
EAC: European Astronaut Centre
ESA: European Space Agency
EVA: Extra-Vehicular Activity
FPGA: Field Programmable Gate Array
GPU: Graphics Processing Unit

HMD: Head Mounted Display
IMU: Inertial Measurement Unit
ISS: International Space Station
LEO: Low Earth Orbit
LBE: Location Based Entertainment
OBT: On-Board Training
PFC: Parabolic Flight Campaign
VR: Virtual Reality
VR-OBT: Virtual Reality On-Board-Training
XR: Umbrella term that covers all "immersive" technologies, including AR, MR, and VR

## 1. Introduction

In recent years, the European Astronaut Centre (EAC) of the European Space Agency (ESA) and the German Aerospace Center (DLR) have significantly advanced the application of novel eXtended Reality (XR) technologies to enhance various aspects of spaceflight operations. XR technologies are now being leveraged to create digital twins of buildings and spacecraft, as well as to revolutionize astronaut training both on the ground and in orbit.

For XR technologies to be effectively utilized in spaceflight, it is essential to validate them under conditions that closely mimic the space environment. This is particularly crucial for the sensors in XR Head-Mounted Displays (HMDs), which can face challenges in microgravity. Although gravitational acceleration is still present in orbit, the conditions of continuous free fall can cause the headset's sensor-fusion algorithms to produce erroneous calculations, requiring adjustments to ensure reliable performance in such microgravity environments.

Simulating weightlessness on Earth is imperative for testing these modifications. One approach involves the use of drop towers, such as the ZARM drop tower in Bremen, Germany, which can create microgravity conditions for 4.74 to 9.3 seconds [1]. However, due to the subjective nature of virtual reality (VR) and the need for user interactions with tracked controllers, drop towers present limitations for comprehensive VR experiments.

An alternative and more effective method to simulate weightlessness is through parabolic flight campaigns (PFCs). During these flights, specially equipped aircraft follow a precise parabolic trajectory to generate periods of microgravity or other reduced gravity conditions, such as lunar or Martian gravity, within the cabin. Depending on the specific flight profile, these reduced gravity phases last between 21 and 30 seconds [2].

The advantage of parabolic flights is that they allow XR devices to be worn and used by participants under simulated microgravity conditions, enabling the assessment of headset tracking accuracy and user interaction with tracked controllers in representative environmental conditions.

This study explores the utilization and testing of XR technology in parabolic flight campaigns to validate its effectiveness for astronaut training. By examining the performance of various XR headsets and motion tracking systems under lunar gravity conditions, this research aims to establish guidelines for integrating XR into future lunar training and preparation activities.

## 2. Related Work

In 2022, the EAC in cooperation with DLR and the French space agency (CNES) launched its first VR experiment on board the International Space Station (ISS) during the Cosmic Kiss mission with ESA Astronaut Matthias Maurer.

VR on-board training (VR-OBT1) utilized a commercial-off-the-shelf (COTS) Oculus Quest Gen 1 headset [3].

Standalone VR headsets, like the Oculus Quest used in this training application, depend on multiple sensors to accurately perceive the environment and calculate the user's position and orientation. However, embedded sensors such as accelerometers and magnetometers, which are part of the IMUs necessary for tracking, face significant challenges in microgravity. These sensors are calibrated for Earth's 1g environment and rely on a stable magnetic field as a reference, conditions that are not matched on the ISS. As a result, this leads to issues such as unwanted drift, jitter, and flicker, which interfere with the accurate display of training content.

Consequently, since COTS devices are not designed to operate in these conditions, the tracking algorithm needs to be adapted to the ISS environment to address issues like unwanted drift, jitter, and flicker, which interfere with the accurate display of training content.

The EAC team therefore developed software to compensate for the expected drift, and the Oculus Quest was tested in a 0g parabolic flight campaign. Cameras mounted to the plane's handrail recorded the outside view during the parabolas, while screen recordings captured the in-headset experience.

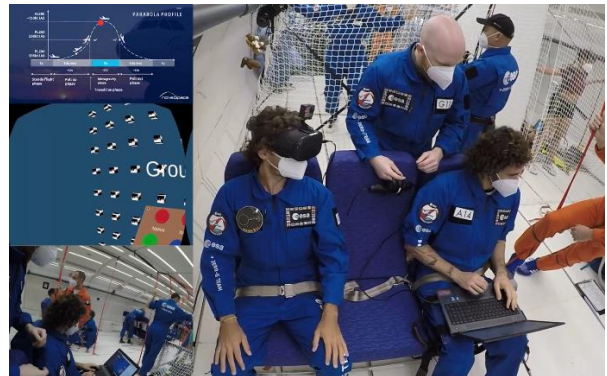

Fig. 1. Synchronized video streams from parabolic flight testing: current flight position (top left), Quest user's view (middle left), and GoPro view from the Quest (bottom left).

However, due to insufficient access to the device's operating system and inability to adjust the tracking algorithm, the implemented drift correction solution had to be performed on application level.

On the ISS, an additional step of capturing the current drift by recording the IMU sensor values just before the execution of the experiment was implemented, which was then applied for a dynamic drift correction and recentering within the training application on application level to counteract drift [4].

Although initially showing rather promising results in the parabolic flight, (the reason for this is likely that the magnetometer was still functioning, resulting in two correct sensor outputs and one erroneous sensor output), after the drift correction algorithm was enabled on-board, the correction and the application did not work as expected. As reported by ESA astronaut Matthias Maurer, the scene drift was not noticeable but instead a very strong flickering and jitter was present.

Therefore, the VR training procedure could not be performed correctly. CNES, who was also using the Oculus Quest 1 for their experiment PILOTE and Immersive Exercise, had described similar issues to us [3].

## 3. Methods/Experiment Design

As previously mentioned, utilizing parabolic flight campaigns is crucial for exploring the potential of XR technology in astronaut training and its application in microgravity conditions for future lunar and Low Earth Orbit (LEO) missions. The EAC recently conducted a series of experiments during a mars and lunar gravity parabolic flight campaign to test the performance and reliability of various XR devices and a motion capture setup in these unique environments.

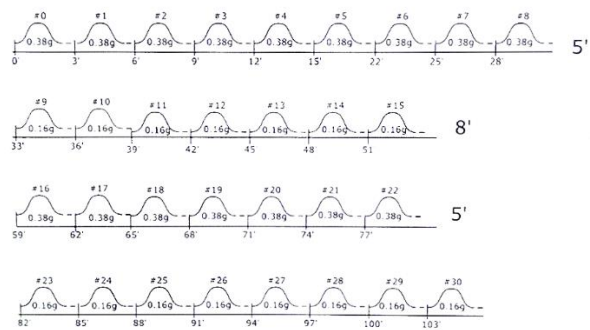

Fig 2. Parabola plan followed during the second flight of the campaign

The experiments involved testing three different VR systems and one motion capture system.

### Varjo-XR3 (XR-MoonGrav+)

The first experiment was situated in the back right of the A310 0g aircraft. Of all experiments in the plane, it had the largest experiment area with around 12m².

The goal of the XR-MoonGrav experiment was to assess whether users scould successfully grasp and carry a tracked object, specifically a weighted box, over a distance of a few meters during a simulated lunar EVA. This required a relatively large experiment area.

The setup included a Varjo XR-3 enterprise VR headset, which was connected by cable to a VR-capable notebook equipped with an i9-12900H CPU and an RTX 3080Ti Mobile GPU. The notebook was mounted on a rack adjacent to the experiment area and ran the Varjo

Base software along with our custom SteamVR application, "XR-MoonGrav."

SteamVR was used as the tracking system, with two second-generation HTC Vive base stations deployed. A third-generation Vive Tracker was attached to the weighted box, enabling its movements and dimensions to be accurately mirrored in the virtual environment.

A precise virtual replica of the box was created and aligned with its real-world counterpart to ensure perfect overlap.

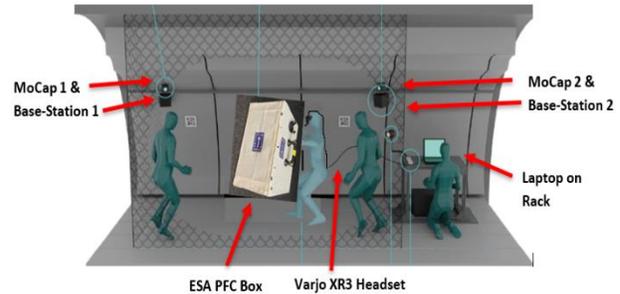

Fig. 3. Diagram of the XR-3 and MoCap setup

When donning the headset, the user was placed in a virtual environment of the lunar south pole. This is a slightly modified and scaled down version of our 64km² recreation of the lunar surface VORTEX, which is primarily used for Concept Operations (ConOps) with ESAs lander Argonaut.

In the exercise run, for the duration of the parabola, the headset was calibrated to a predefined position on the rack and then the box had to be carried in VR to a marked location on the virtual ground, this location would also match the box that was placed on the aircraft cabin floor. If there was any error / discrepancy in the tracking of the user, it would be visible after the user had placed the box correctly virtually.

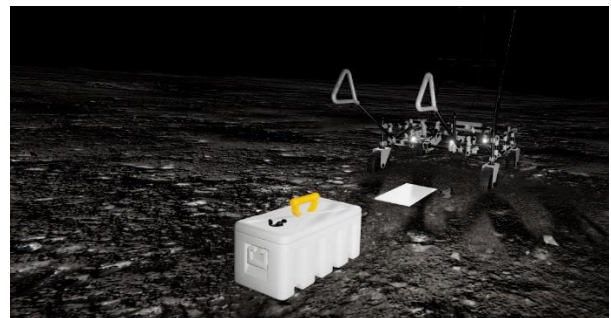

Fig. 4. Unreal Engine 5 "XRMoonGrav" application

After each parabola and its associated hypergravity phase, the user removed the headset, and the application was reset during the interval between parabolas. The goal was to complete at least 10 runs in both lunar and Mars gravity conditions over the course of three flight days.

In addition to the XR-MoonGrav experiment, we also tested the Varjo XR-3's other tracking modes, specifically the inside-out tracking mode and a specialized 0g tracking mode that required markers to be placed on the wall.

### Azure Kinect (MoCap)

In parallel to the XR-MoonGrav experiment, a motion capture experiment was conducted within the same experimental area. This setup involved two Microsoft Azure Kinect cameras, positioned opposite each other, with one camera rotated 180 degrees to face the other. Both cameras were connected to the same notebook as the Varjo XR-3 headset, although the two experiments were conducted separately.

The use of two cameras was essential to reduce potential occlusion and enhance the overall accuracy of the motion capture. To achieve synchronized and precise captures, the cameras were calibrated to work in tandem. This synchronization was accomplished by connecting the cameras with a 3.5mm standard audio cable. Additionally, the camera positioned farther from the notebook was connected via a special fiber-optic cable, ensuring high data throughput and stable power delivery.

Motion capture data was processed using the Brekel Body v3 full-body capture software running on the laptop [5]. Calibration of the two Kinect sensors involved performing a series of calibration movements, which enabled the software to generate an adjusted point cloud and accurately capture the user's joint movements.

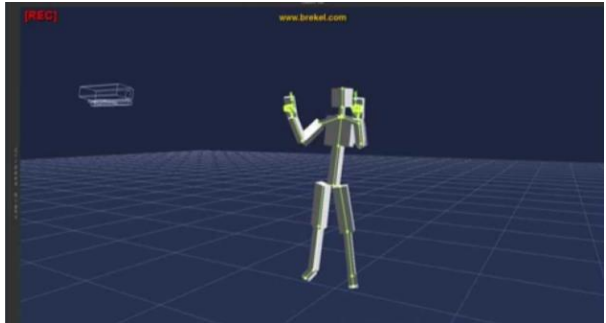
Fig. 5. 3D View of the "Brekel Body v3" capture software

The experimental procedure was as follows: The system was calibrated on the ground prior to the flight. During the parabolic flight, recordings were initiated when stable low-g conditions had been achieved. The participant performed moon jumps and hops in one direction, followed by a return to the starting point. After each sequence of movements, the recording was stopped. This procedure was repeated multiple times during 0.16g and 0.38g phases throughout the flight.

Recordings were saved in FBX format to facilitate detailed post-processing and analysis. The setup was closely aligned with the Varjo experiment, as it utilized the same experimental area and computer.

### HTC VIVE Focus 3

The second experiment, positioned across the aisle from the XR-MoonGrav setup but at the same section of the aircraft, involved an HTC VIVE Focus 3 standalone VR headset, which, for evaluation purposes, was also connected to a notebook of the same model. Like the XR-MoonGrav experiment, this setup utilized a custom-built Unreal Engine 4 application, where users were required to perform motions with the right VR motion controller to hit moving targets within the virtual environment.

We had specifically chosen to test the HTC VIVE Focus 3 after discovering its VR Simulator mode [6], which we believed could significantly enhance tracking stability in microgravity environments, making it an ideal candidate for our parabolic flight experiments.

To adapt the Focus 3 for the unique conditions of the parabolic flight, we utilized this VR Simulator mode. This mode is particularly advantageous in environments where traditional tracking methods may not be reliable, such as microgravity or other dynamic motion conditions. By using an external reference point rather than relying solely on its internal IMU and cameras, the Focus 3 can maintain accurate tracking even in challenging environments.

For this experiment, the left controller was designated as the tracking anchor and securely mounted on a metal pole, which was fixed to the aircraft floor. This setup provided a stable reference point, allowing the headset to maintain accurate tracking even as the aircraft transitioned through different gravity phases.

The tracking method used here is akin to motion cancellation, where the movement detected by the headset is compensated by the stationary position of the fixed controller. This technique is commonly employed in location-based entertainment (LBE) venues and motion simulators to maintain a stable virtual environment [6].

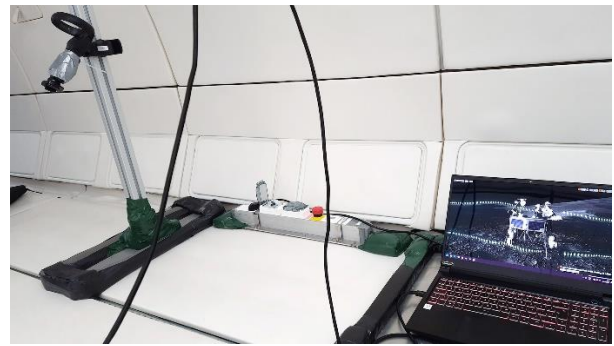
Fig. 6. The setup involved securely fastening the left hand-controller to a metal pole, using it as a stable tracking anchor

## Custom HMD

The third system was a custom-built standalone HMD specifically designed in-house to offer reliable 6DoF tracking in microgravity environments, by relying on optical tracking only. To validate the accuracy of this tracking system, passive markers and an OptiTrack V120:Trio were employed to provide ground truth data.

This custom HMD ran an Unreal Engine 4 application. Meanwhile, the ground truth tracking data from the OptiTrack Duo V2 was captured and processed on a separate notebook equipped with an Intel Core i7-12700H processor and an RTX 3070 Mobile GPU.

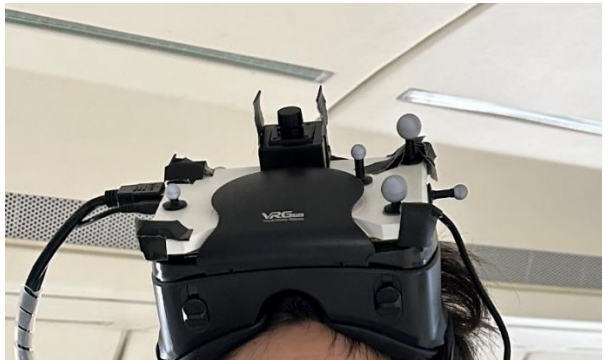

Fig. 7. Custom VR HMD "Spacetacles" with optical markers attached

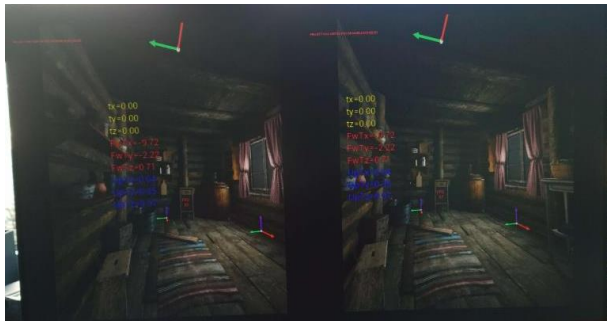

Fig 8. The Unreal Engine 4 application display output

The experiment aimed to demonstrate that the bespoke tracking system performs consistently under different gravity levels. A conservative approach was followed, from isolated component tests to freeform head movements.

Testing started with commissioning, which included isolated tests of the OptiTrack system, stereoscopic display, and computer under various gravities. Following the commissioning, the bespoke tracking system was tested under continuous performance monitoring. Multiple parabolas, including phases of hyper-gravity, were used at each stage of testing:

1. 0-reference: the head kept still. The goal here is to assess any false motion detected by the system.

2. Slow and small oscillating roll motion. This step is often the easiest motion for a tracking system as it doesn't change the sensor input as much as the other rotations.
3. Slow and small oscillating tilt motion
4. Slow and small oscillating pan motion
5. +/-90degree roll oscillations. This step serves to detect if or where tracking fails. The first parabola is about finding if a breaking point exists, and the next parabolas are for assessing the stability of the system when staying away from the breaking point.
6. Maximum tilt oscillations. Same as above, for the tilt movement.
7. Maximum pan oscillations. Same as above, for the pan movement.
8. Freeform tilt and pan movement, while staying within the working envelope.
9. Freeform rotation on all axes in the working envelope. This step is where the motions of the test operators are expected to be as organic as possible, in contrast to the robotic ones performed previously.

A control test was also conducted on the ground. After each flight, participants provided feedback via questionnaire, and the testing methodology was adjusted accordingly.

## 4. Results

### Varjo XR-3

The Varjo XR-3 encountered significant challenges during our series of parabolic flights, particularly when used with SteamVR. During the first flight, under hypergravity and microgravity conditions, the headset consistently lost tracking, and the virtual ground appeared heavily offset. These issues were traced back to the base stations, which rely on IMUs to stabilize tracking between each other.

During steady 1-g flight conditions, the Varjo XR-3's tracking capabilities with SteamVR were mostly satisfactory, though some minor issues were observed, such as a slight jitter and a perceptible tilt in the virtual reality environment. However, more significant challenges arose during lunar-g and hyper-g conditions, where the SteamVR tracking system failed to initialize, rendering the headset, controllers, and trackers inactive with no positional or rotational tracking. This behavior was similar to the issues we observed with the Oculus Quest during parabolic flights in 2021, where the reliance on built-in IMUs led to significant tracking failures in microgravity environments.

To address these problems, we modified the base station driver before the second flight, disabling the IMU functionality. This modification led to immediate improvements; the base station logs no longer showed the 'unlikely gravity - resetting' error, and SteamVR initialized consistently across all gravity conditions, including steady flight, hyper-g, and lunar/mars-g scenarios.

However, despite these improvements, persistent jitter and flicker in the VR scene were observed both on the ground and during flight, manifesting in the SteamVR environment and within the Unreal Engine application. While this issue appeared after the driver modification, it could also be influenced by other factors, such as aircraft-related vibrations or structural conditions.

While the hand controllers, headset, and trackers maintained mostly accurate positions and rotations in the virtual environment, the flickering presented a significant challenge during the experiments.

Varjo had provided us with a 0g tracking plugin specifically designed to mitigate these issues in microgravity conditions. However, despite our efforts, we were unable to get this plugin to function correctly during the third and final flight day, leaving the Varjo XR-3 affected by the same challenges we encountered earlier.

In exploring the Varjo XR-3's Mixed Reality Mode under two distinct tracking conditions, SteamVR mode and Inside-Out tracking mode, we found varying degrees of performance. In Inside-Out tracking mode, during steady flight, the virtual elements displayed commendable steadiness initially. However, this stability was short-lived; within seconds, the windows began to drift noticeably from the calibrated center, sometimes even flipping upside down or drifting out of view, making it difficult to maintain orientation.

In contrast, using the SteamVR mode with the modified driver, the virtual scene's positional integrity was mostly preserved, but the persistent jitter and flicker continued to affect the overall experience. Despite these disruptions, the tracking remained operational, allowing us to execute our intended experiments. However, the visual disturbances posed a challenge by impairing the test subject's hand-eye coordination and orientation during the trials.

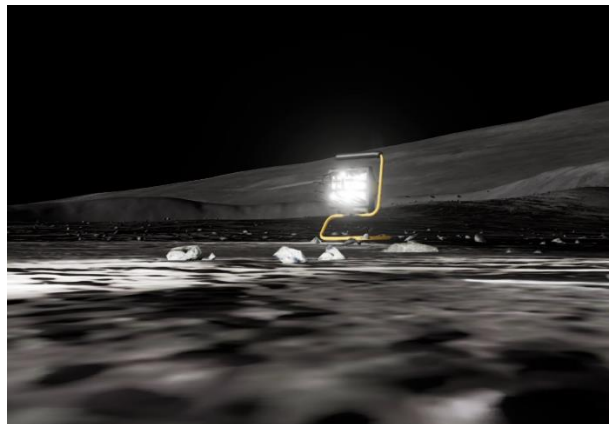

Fig. 9. SteamVR tracking failed to initialize

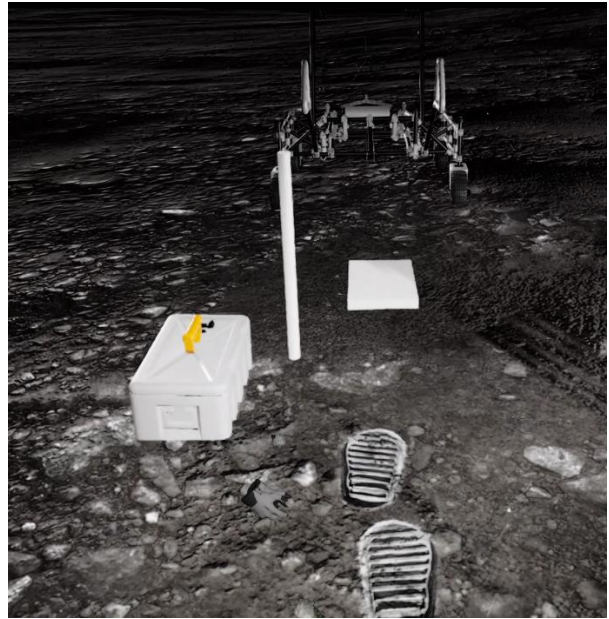

Fig. 10. Correctly initialized SteamVR tracking

### Azure Kinect

The Azure Kinect performed reasonably well after proper calibration before each flight. The aforementioned dual-camera setup, positioning two Microsoft Azure Kinects to face each other at an approximately 180-degree angle, maximized the coverage of the desired tracking space. The cameras were mounted on the plane's handrail at head level and spaced about 4.5 meters apart, providing a substantial tracking area. To prevent capturing unnecessary movement, especially from passengers and security staff traversing the middle aisle, the cameras were tilted slightly inwards.

As long as only one person was within the capture area, tracking performance remained consistent, allowing us to successfully perform over 50 lunar-g captures. However, post-processing of the captured data in fbx format is essential to trim unnecessary footage outside the desired timeframe. Additionally, there were occasional instances where individual joints stopped tracking accurately and only followed the movement of the parent joint.

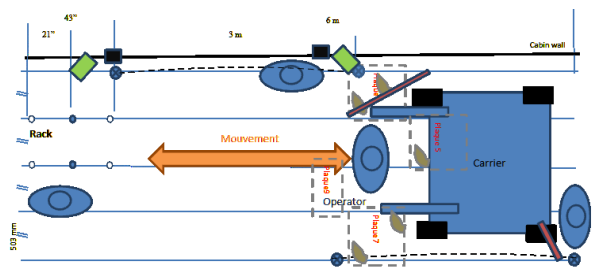

Fig. 11. XR-3 and MoCap layout in the A310 PFC plane

 

The tracking calibration procedure was rigorously performed before every flight to maintain the integrity and accuracy of our data. Overall, the motion capture sessions using the Brekel Body v3 software were largely successful, demonstrating the viability of the dual Azure Kinect setup for capturing motion.

### HTC VIVE Focus 3

The performance of the HTC VIVE Focus 3 during our parabolic flight experiments was excellent overall. The Simulator VR mode functioned as intended, providing stable tracking throughout the majority of the flight manoeuvres. However, we observed that the left controller needs to be securely fixed to a surface, such as a wall, and must remain unobstructed to maintain accurate positional tracking. Any occlusion led to a loss of tracking, highlighting the importance of a clear line of sight for optimal performance.

A notable issue arose with the aircraft's cabin LED light strips. When the LED light directly above the experiment area was turned on, the Focus 3 experienced significant tracking failures. This interference likely stemmed from the infrared spectrum overlap between the LED lights and the headset's tracking system. The LED lights, operating at a flicker frequency of 366 Hz and with brightness levels between 1.4 and 1.9 klux, caused both the controller and headset orientation tracking to be compromised, leading to a complete loss of tracking capability.

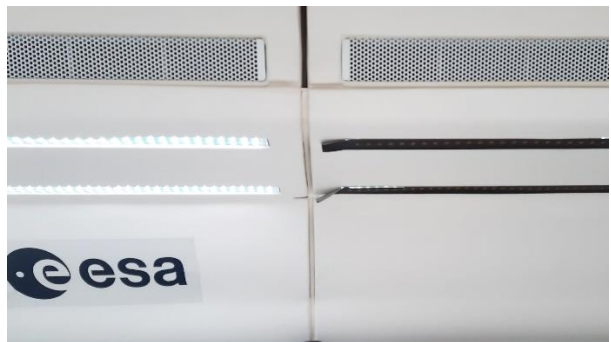

Fig. 12. The cabin LED lights had to be taped over during the experiment to prevent tracking issues

Despite these challenges, the HTC VIVE Focus 3 performed reliably across most parabolas, with only minor jitter detected when turning too far left or right, causing the headset to lose accurate tracking of the controller, or when the controller was occluded.

The motion cancellation mode worked effectively, with the headset exhibiting almost no jitter under these conditions. However, the issue with the cabin lighting necessitates further investigation, particularly for future experiments in similar environments like the Columbus module on the ISS, where lighting conditions must be carefully controlled to ensure seamless VR operation.

The application, streamed to the Focus 3 via USB cable, operated smoothly, and we expect similar performance if the application were to run directly on the headset.

### Custom HMD

The custom HMD performed admirably with no tracking malfunctions throughout the flight campaign, and no discernible breaking point in roll motion. The testing methodology allowed operators sufficient time to observe potential issues, leading to several improvements:

- The initial test plan lacked translation on the X-Y-X axes. This was added during the third flight, adapting steps 5-9 based on feedback.
- Sudden fast movements were not explicitly tested.
- An additional measurement was implemented in the third flight, adding a moving target for each rotation axis. Operators tracked the target with their gaze while the software recorded precision.
- The tracking system exhibited higher latency compared to other COTS headsets, with approximately 400ms of "Motion-to-Photon" latency. This latency is uncomfortable for real-world use, and there were no precise tools to measure it. The combination of OptiTrack and the bespoke system wasn't synchronized with enough temporal precision.
- Increasing the number of headset operators is recommended for higher technological readiness, especially for real-world use cases, although this is not essential for assessing gravity-level independence.

Nevertheless, the gravity level in the plane had no discernible impact on the tracking system's performance, providing empirical evidence of gravity independence.

## 5. Discussion

The testing of three different VR headsets provided valuable insights into the tracking process and gave us considerable experience with these devices under reduced gravity conditions. Additionally, we have gained a clearer understanding of the limitations of each device, enabling us to develop more effective usage guidelines - particularly for the HTC VIVE Focus 3 headset's potential application on the ISS.

Following the successful tests during the PFC, the Focus 3 was selected as the headset of choice for VR experiments and activities on board the ISS. The adapted tracking algorithm delivered impressive results during the parabolic flights, reinforcing our belief that this approach will be the most viable solution in the near term, until a headset specifically designed for microgravity environments can be space-qualified and deployed.

The Focus 3 performed admirably in microgravity, making it a suitable tool for our VR experiments on the ISS. While it necessitates sacrificing the left controller for interactions, this trade-off is balanced by the fact that no software modifications are required. Once the controller is securely mounted to a stable structure, such as a wall, and detected by the headset, the system functions effectively. However, setting up the VR system can be time-consuming and cumbersome, as it requires calibrating the headset for both direction and floor height after properly securing the controller to a stationary surface.

Despite the promising results with the Focus 3, we recognize the importance of continuing development on our custom 0g HMD. This is particularly crucial given the stringent requirements for safety, security, and reliability in space environments. For instance, modifications to the headset and controller batteries, as well as regular safety extensions, are necessary to meet the rigorous standards of space missions.

By addressing these requirements, we aim to create a headset that not only meets the technical challenges of microgravity but also adheres to the strict safety protocols required for deployment on the ISS.

Based on our findings, we have established several key guidelines for improving VR headset performance in microgravity. First, IMUs in VR headsets can be unreliable in these conditions, leading to tracking errors, so alternative tracking methods should be considered. In our case, modified tracking solutions, such as fixed reference points with software adjustments like the VR simulator mode by HTC or using optical only tracking, have proven effective in improving accuracy.

It is also important to mitigate environmental factors, such as LED lighting interference and aircraft vibrations, which can disrupt tracking. Lastly, we recommend simplifying test procedures, as cognitive capacity can decrease rapidly in environments like microgravity.

## 6. Conclusions/Future Work

After the positive results in the PFC, additional tests were performed during the Axiom 3 mission, which successfully replicated our findings and further validated the use of the HTC VIVE Focus 3 headset for VR activities and training on board the ISS. The vision for the future is to provide astronauts with an engaging and immersive way to conduct just-in-time training through XR devices.

As XR technology continues to evolve, with advancements in both hardware and software, we anticipate a growing demand for testing these devices in parabolic flight campaigns. This will be crucial as we expand from pure VR training to incorporating Mixed Reality (MR) and Augmented Reality (AR) devices, such as the Meta Quest 3 and Apple's Vision Pro, into astronaut training.

Our goal is to continually push the boundaries of what these technologies can offer, not only in enhancing training for astronauts but also in providing recreational experiences that contribute to their well-being during long missions [7].

In parallel, we remain committed to refining our custom-built VR headset for microgravity. There is significant potential to reduce latency by integrating custom processors, such as Field Programmable Gate Arrays (FPGAs), to handle the tracking algorithms more efficiently. Additionally, implementing techniques such as Asynchronous Reprojection or Asynchronous Time Warp, commonly used in commercial headsets, could further enhance performance. These techniques generate intermediate frames when the application struggles to maintain the native refresh rate, reducing lag, tearing, and the motion-to-photon latency, thereby improving the overall user experience [8].

As we continue our research and development efforts, we look forward to further advancing the capabilities of XR technology in space, ensuring that astronauts have access to the most cutting-edge tools for training and recreation in the unique environment of microgravity.

## Acknowledgements

The authors would like to sincerely thank Novespace and the 81st ESA parabolic flight team for their crucial support, enabling extensive testing during the campaign.

We are also grateful to DLR and ESA for their collaboration and the exceptional cooperation of the teams at the European Astronaut Centre (EAC), which made this project possible.

Finally, we appreciate everyone who contributed valuable knowledge from previous VR experiments in hypogravity, guiding our work.